\documentclass[conference]{IEEEtran}
\IEEEoverridecommandlockouts

\usepackage{cite}
\usepackage{amsmath,amssymb,amsfonts}
\usepackage{graphicx}
\usepackage{textcomp}
\usepackage{xcolor}
\usepackage{booktabs}
\usepackage{multirow}
\usepackage{array}
\usepackage{url}
\usepackage{eso-pic}
\usepackage{tikz}
\usepackage{pgfplots}
\pgfplotsset{compat=1.17}
\usetikzlibrary{positioning,shapes.geometric,arrows.meta}

\def\BibTeX{{\rm B\kern-.05em{\sc i\kern-.025em b}\kern-.08em
    T\kern-.1667em\lower.7ex\hbox{E}\kern-.125emX}}

\begin{document}

\AddToShipoutPictureBG*{%
  \AtPageLowerLeft{%
    \raisebox{5mm}{%
      \makebox[\paperwidth]{%
        \parbox{0.86\paperwidth}{\centering\scriptsize
          Published in the 2026 IEEE International Conference on LLM-Aided
          Design (ICLAD). Final version: https://doi.org/10.1109/ICLAD71857.2026.00026\\
          \textcopyright~2026 IEEE. Personal use of this material is permitted.
          Permission from IEEE must be obtained for all other uses, in any
          current or future media, including reprinting/republishing this
          material for advertising or promotional purposes, creating new
          collective works, for resale or redistribution to servers or lists,
          or reuse of any copyrighted component of this work in other works.}}}}}

\title{Inference-Time Scaling in Agentic Hardware Verification Workflows:\\
A Multi-Strategy Characterization Across Protocol Complexity}

\author{\IEEEauthorblockN{Vikash Kumar}
\IEEEauthorblockA{\textit{Arm}\\
Austin, Texas, USA\\
vikash.kumar4@arm.com}}

\maketitle

\begin{abstract}
Inference-time scaling, allocating more compute during generation
rather than pre-training, has demonstrated dramatic quality
improvements for LLM reasoning tasks. Its behaviour in structured
hardware verification code generation, where outputs must satisfy
compiler correctness, protocol semantics, and functional coverage
constraints simultaneously, is less understood. We present a
systematic, simulator-grounded study of inference-time compute
scaling for agentic UVM testbench synthesis targeting AMBA
protocol bridges. Four inference strategies---zero-shot
(\$0.04), raw-feedback iterative (\$0.81), structured inner-loop
(\$0.95), and full dual-loop agentic synthesis
(\$1.06)---are evaluated across four AMBA targets (Protocol
Complexity Index 11.8--22.6) using the Protocol Verification
Completeness (PVC) metric, functional coverage, and a six-class
UVM error taxonomy (E1--E6). Our central finding is that
\emph{structured diagnostic feedback acts as an inference-time
compute multiplier}: the structured inner loop yields roughly
$15\times$ more PVC per inference dollar than raw EDA-tool
feedback (marginal efficiency, S1$\to$S2 vs.\ S0$\to$S1). The
mechanism is fault targeting, not token reduction: raw compiler
feedback drives the testbench to compile but cannot surface
functional protocol faults that leave the code syntactically
valid, whereas structured E1--E6 bundles diagnose and resolve
them. All outputs are judged by a cycle-accurate Mentor Questa
toolchain rather than by model self-assessment; the
dual-loop strategy reaches PVC\,$\geq$\,0.95 with zero scoreboard
errors on every target, and the generated environments surfaced
a latent RTL bug in two of the four reference designs. These
results establish a principled compute-quality trade-off curve
for LLM-aided verification environment generation and motivate
inference-budget-aware deployment strategies for industrial EDA
workflows.
\end{abstract}


\section{Introduction}
\label{sec:intro}

Functional verification consumes 60--70\% of SoC design
effort~\cite{foster2022wilson}, a fraction that grows as
chiplet-based integration multiplies the number of protocol
bridge instances requiring independent testbenches. A protocol
bridge UVM (Universal Verification Methodology) environment for
a moderate interface (AHB, APB, AXI-Lite) runs to 3{,}000--8{,}000
lines of SystemVerilog and demands 40--100 engineering hours of
bring-up effort. Unlike RTL generation, where a single synthesis
call can yield a functionally complete
module~\cite{blocklove2025autochip}, UVM testbench generation
requires simultaneous satisfaction of three orthogonal
constraints: \emph{structural correctness} (UVM class hierarchy
and TLM connectivity), \emph{protocol semantics} (LTL-encoded
handshake ordering), and \emph{coverage completeness}
(intent-driven cross-product covergroups). Failing any one
constraint produces a testbench that either does not compile,
does not model the protocol correctly, or misses critical corner
cases.

Recent work has shown that \emph{inference-time
scaling}, allocating more compute during generation through
iterative refinement, search, or feedback loops, can substitute
for model scale and dramatically improve output quality on
challenging tasks~\cite{snell2025scaling}. Yet this paradigm has
not been systematically applied or characterised in hardware
verification code generation, where the ``verifier'' is a
deterministic EDA compiler and simulation engine, the
``difficulty signal'' is a protocol-specific error taxonomy, and
the ``prompt'' is a structured Verification Manifest encoding
interface topology and coverage intent.

\textbf{Scope of claims.} This paper studies UVM testbench
synthesis for AMBA protocol bridges (APB, AHB, AXI-Lite) on
four reference DUTs in the PCI~11.8--22.6 range, with Claude
Opus~4.8 as the synthesis model and Mentor Questa as the
verification toolchain. We do not claim generalisation to
non-AMBA protocols, designs with multiple asynchronous
clock-domain crossings, formal verification flows, RTL repair,
or non-UVM methodologies; these are deferred to future work.

We ask: \emph{given a fixed inference budget for verification
workflow generation, how should that budget be allocated across
feedback strategy and iteration depth?} Our contributions are:

\begin{itemize}
\item A four-strategy inference-time compute taxonomy for
agentic UVM synthesis (S0--S3), spanning zero-shot to full
dual-loop synthesis, enabling systematic cost-quality
characterisation (\S\ref{sec:strategies}).

\item Empirical evidence that structured diagnostic feedback
(the E1--E6 UVM error taxonomy) is a \emph{compute multiplier}:
the structured inner loop yields $\sim\!15\times$ more PVC per
inference dollar than raw EDA-tool feedback, with the mechanism
identified as functional-fault targeting rather than token
reduction (\S\ref{sec:results}).

\item A simulator-grounded evaluation methodology in which every
generated testbench is judged by a real EDA toolchain (Mentor
Questa compile, simulation, scoreboard, and functional coverage)
rather than by model self-assessment; the generated environments
additionally surfaced a latent RTL bug in two of the four
reference designs (\S\ref{sec:bugfind}).
\end{itemize}

\section{Background and Related Work}
\label{sec:related}

\textbf{Inference-Time Scaling.}
Snell~et~al.~\cite{snell2025scaling} showed that optimal allocation
of test-time compute, through process-reward-model-guided search
or iterative self-refinement, outperforms pre-training scaling
for reasoning-intensive tasks. Scaling effectiveness depends
critically on task difficulty: easy tasks benefit from iterative
revision, while hard tasks require parallel search. We instantiate
this framework for UVM synthesis, where the verifier is a
deterministic EDA toolchain providing exact, not probabilistic,
correctness signals.

\textbf{LLM-Aided Hardware Code Generation.}
AutoChip~\cite{blocklove2025autochip} pioneered EDA-tool-feedback-driven
Verilog RTL repair, achieving 24.2\% improvement in functional
correctness over zero-shot generation; raw compiler output
is most beneficial for the largest models, with smaller models
deriving less benefit from unstructured diagnostic streams.
RTLFixer~\cite{tsai2024rtlfixer} addresses RTL syntax repair via
LLM feedback but generates no verification environments.

\textbf{LLM-Based Testbench Generation.}
CorrectBench~\cite{qiu2025correctbench} introduced functional
self-correction for Verilog testbench generation, achieving
70.13\% pass rate on HDLBits. UVM$^2$~\cite{ye2025uvm2} demonstrated
multi-agent UVM testbench generation with coverage feedback,
86.67\% generation success on nine RTL designs. Both accept
natural-language input and target single-interface designs
without protocol-semantic coverage. Concurrent
work~\cite{llm4cov2026} demonstrated execution-grounded agentic
training enabling a 4B model to exceed frontier model
performance on coverage-driven testbench generation on HDLBits-class
circuits. None of these characterise the inference-time
cost-quality trade-off space or scaling across protocols of
varying complexity. Complementary ML approaches target
downstream UVM verification productivity, e.g., flaky-test
detection in regression runs~\cite{kumar2026flaky}.

\textbf{Positioning.}
Our work differs from prior art in three ways:
(i)~we study \emph{how} to allocate inference budget, not just
\emph{whether} feedback helps;
(ii)~we target full UVM environments for protocol bridge IP
(3{,}000--8{,}000 lines), not toy modules or assertion
generation~\cite{bai2025assertionforge}; and
(iii)~we identify structured diagnostic feedback as a compute
multiplier with quantifiable per-dollar efficiency, validated
against a real EDA toolchain.

\section{Inference Strategy Taxonomy}
\label{sec:strategies}

We define four strategies (S0--S3) as an ordered sequence of
increasing compute investment and feedback sophistication. All
share the same input: a \emph{Verification Manifest}, a
structured JSON document encoding interface topology, protocol
rules (LTL properties), and coverage intents. The manifest is
fixed across strategies to isolate the effect of the feedback
mechanism.

\textbf{S0: Zero-Shot Synthesis.} A single LLM call generates
all UVM components in strict dependency order: interface,
transaction, sequence, driver, monitor, scoreboard, coverage,
environment, and test. No compiler feedback. Cost: $\sim$\$0.04.

\textbf{S1: Raw-Feedback Iterative.} After each component is
generated, the EDA compiler is invoked and its full, unstructured
diagnostic output is appended to the LLM context. The LLM
repairs the component and reinvokes the compiler. Six maximum
inner iterations per component, the closest analogue to
AutoChip-style raw injection~\cite{blocklove2025autochip} applied
to UVM. Cost: $\sim$\$0.81.

\textbf{S2: Structured Inner-Loop Only.} Identical to S1 except
that compiler diagnostics are pre-processed by a UVM Error Parser
before LLM injection. The parser categorises each diagnostic into
one of six classes (E1--E6, \S\ref{sec:taxonomy}) and injects a
structured bundle: error class, affected component, code
context, violated LTL property, and fix template. No outer
coverage loop. Cost: $\sim$\$0.95.

\textbf{S3: Full Dual-Loop Agentic Synthesis.} Extends S2 with
an outer coverage feedback loop. Once a compile-clean testbench
is produced, it is executed against a golden reference VIP, and
an intent-gap analyser compares achieved coverage against the
Manifest's \texttt{coverage\_intents[]}. Uncovered intent bins
are re-injected into the Manifest, triggering re-synthesis of
affected sequence and covergroup components. The outer loop
drives functional coverage to closure. Cost: $\sim$\$1.06.

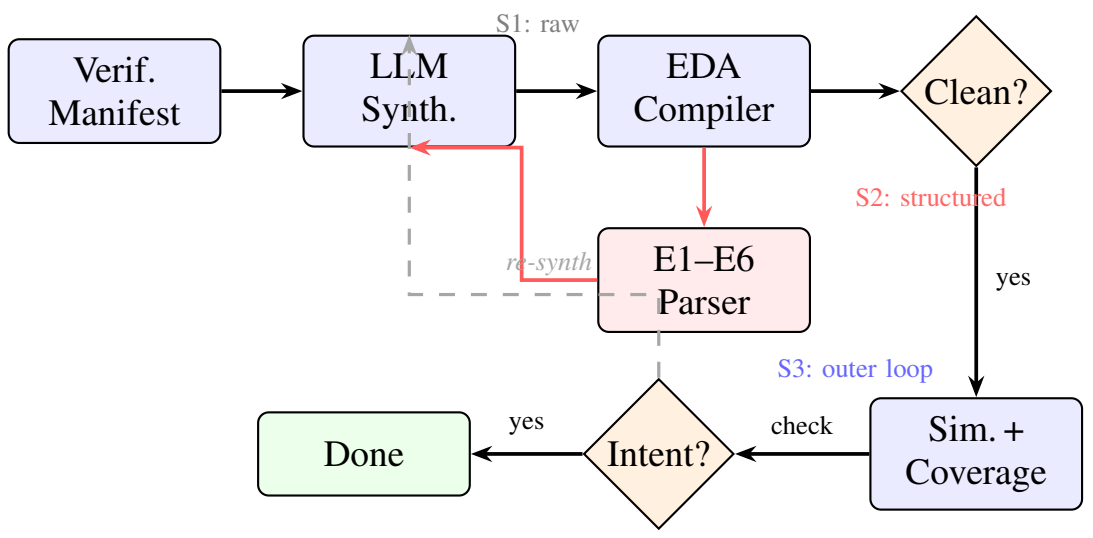
\begin{figure}[t]
\centering
\begin{tikzpicture}[
  font=\scriptsize,
  box/.style={rectangle,draw,rounded corners=2pt,
              minimum width=1.4cm,minimum height=0.55cm,
              align=center, fill=blue!8},
  decision/.style={diamond,draw,fill=orange!12,
                   minimum width=0.9cm,minimum height=0.55cm,
                   align=center, inner sep=1pt},
  arr/.style={-{Stealth[length=4pt]},line width=0.7pt},
  redarr/.style={-{Stealth[length=4pt]},line width=0.7pt,red!65},
  dasharr/.style={-{Stealth[length=4pt]},line width=0.6pt,
                  dashed,gray!70},
]
\node[box] (manifest) at (0,0)    {Verif.\\Manifest};
\node[box] (llm)      at (1.95,0) {LLM\\Synth.};
\node[box] (vcs)      at (3.9,0)  {EDA\\Compiler};
\node[decision](ok)   at (5.7,0)  {Clean?};

\node[box,fill=red!8] (parser) at (3.9,-1.25) {E1--E6\\Parser};

\node[box] (sim)      at (5.7,-2.4) {Sim.\,+\\Coverage};
\node[decision](cov)  at (3.6,-2.4) {Intent?};
\node[box,fill=green!8] (done) at (1.65,-2.4) {Done};

\draw[arr] (manifest) -- (llm);
\draw[arr] (llm)      -- (vcs);
\draw[arr] (vcs)      -- (ok);
\draw[arr] (ok)       -- node[right,font=\tiny]{yes} (sim);

\draw[redarr] (vcs.south) -- (parser.north);
\draw[redarr] (parser.west) -- ++(-0.5,0) |- (llm.south);

\draw[arr] (sim)  -- node[above,font=\tiny]{check} (cov);
\draw[arr] (cov)  -- node[above,font=\tiny]{yes} (done);
\draw[dasharr] (cov.north) -- ++(0,0.55) -|
               node[above,pos=0.22,font=\tiny]{\textit{re-synth}}
               (llm.north);

\node[font=\tiny,gray]    at (2.8, 0.45) {S1: raw};
\node[font=\tiny,red!60]  at (5.4,-0.70) {S2: structured};
\node[font=\tiny,blue!60] at (4.9,-1.85) {S3: outer loop};
\end{tikzpicture}
\caption{S3 dual-loop architecture. The inner loop (S1/S2)
iterates compiler feedback per component; S2 replaces raw
compiler output with structured E1--E6 bundles (red path). The
outer loop (S3 only, dashed) re-synthesises components with
unmet coverage intents. S0 omits both loops.}
\label{fig:architecture}
\end{figure}

Table~\ref{tab:strategies} summarises the strategies and
Fig.~\ref{fig:architecture} illustrates the S3 architecture;
S0--S2 are strict subsets.

\begin{table}[t]
\caption{Inference Strategy Summary}
\label{tab:strategies}
\centering
\small
\begin{tabular}{@{}lcccc@{}}
\toprule
Strategy & Inner Loop & Structured & Outer Loop & Cost \\
\midrule
S0 Zero-shot     & ---     & No  & No  & \$0.04  \\
S1 Raw-feedback  & Raw     & No  & No  & \$0.81  \\
S2 Struct.-inner & Struct. & Yes & No  & \$0.95  \\
S3 Full dual     & Struct. & Yes & Yes & \$1.06  \\
\bottomrule
\end{tabular}
\end{table}

\section{Platform and Metrics}
\label{sec:platform}

\subsection{Agentic Synthesis Platform}
All strategies share a common agentic Python
framework~\cite{kumar2026isvlsi,kumar2026vega,kumar2026metrics}. The LLM
Synthesis Engine generates UVM components in strict dependency
order; each LLM call receives the relevant Manifest fragment,
LTL protocol properties, a structural template, and up to three
exemplar implementations.
Prompt context density of 45--72\% consistently maximises PVC
in our experiments; below 45\% structural context is omitted
(E1--E3 errors rise), and above 72\% E4--E5 protocol faults rise,
likely from attention spread over long contexts.

\subsection{UVM Error Taxonomy}
\label{sec:taxonomy}
\begin{figure*}[t]
\centering
\begin{tikzpicture}[font=\small]
\node[draw, fill=red!5, rounded corners=3pt,
      minimum width=7.5cm, minimum height=3.7cm,
      anchor=north west] (rawbox) at (0,0) {};
\node[anchor=north west, font=\bfseries\small, text=red!70]
  at (0.2,-0.2) {S1 --- Raw Compiler Diagnostic Injection};
\node[anchor=north west, font=\ttfamily\scriptsize,
      text=black!80, align=left]
  at (0.25,-0.55) {%
Error-[ICPD] Illegal combination of drivers\\
~~"apb\_driver.sv", line 47: variable 'psel'\\
~~driven by continuous assignment on line 31,\\
~~and procedurally on line 47.\\
Error-[ELAB] Elaboration failed (line 52)\\
~~Signal pready not in sensitivity list\\
Error-[SV-LCM] Signal already connected\\
~~"apb\_monitor.sv", line 18\\
~~\ldots [11 more errors suppressed]%
  };
\node[anchor=north, font=\scriptsize\itshape, text=red!60]
  at (3.75,-3.85) {$\sim$2{,}760 tokens per E4 resolved};

\draw[-{Stealth[length=6pt]}, line width=1.8pt, blue!60]
  (7.8,-1.85) -- node[above, font=\scriptsize\bfseries,
  text=blue!70, align=center] {Structured\\Parser} (8.6,-1.85);

\node[draw, fill=blue!5, rounded corners=3pt,
      minimum width=7.5cm, minimum height=3.7cm,
      anchor=north west] (strbox) at (8.8,0) {};
\node[anchor=north west, font=\bfseries\small, text=blue!70]
  at (9.0,-0.2) {S2 --- Structured E4 Bundle};
\node[anchor=north west, font=\scriptsize, align=left,
      text=black!85]
  at (9.05,-0.6) {%
\textbf{class:} E4 --- Temporal Protocol Fault\\
\textbf{component:} apb\_driver.sv \quad \textbf{signal:} psel\\
\textbf{violation:} LTL[P2]: PSEL must assert\\
\hphantom{\textbf{violation:}~}before PENABLE (APB \S3.1)\\
\textbf{context:} lines 44--49 (driver.sv)\\
\hphantom{\textbf{context:}~}\texttt{always\_ff @(posedge clk)}\\
\hphantom{\textbf{context:}~}\texttt{~~psel <= ...~// conflict}\\
\textbf{fix\_template:} clocking block to seq.\\
\hphantom{\textbf{fix\_template:}~}PSEL $\rightarrow$ PENABLE\\
\textbf{ltl\_property:} \texttt{P2\_psel\_before\_penable}%
  };
\node[anchor=north, font=\scriptsize\itshape, text=blue!60]
  at (12.55,-3.85) {$\sim$990 tokens per E4 resolved};
\end{tikzpicture}
\caption{Contrast between S1 raw diagnostic injection and S2
structured E4 bundle injection for the same APB driver temporal
fault. Raw injection forces the model to locate the relevant
protocol context within an unstructured multi-error stream; for
a temporal fault the compiler may not report the functional
violation at all. The structured bundle pre-identifies error
class, violated LTL property, code context, and fix template,
directing inference compute precisely at the failing constraint.
This fault targeting is the mechanism behind the structured-feedback
efficiency advantage (\S\ref{sec:tokens}).}
\label{fig:contrast}
\end{figure*}
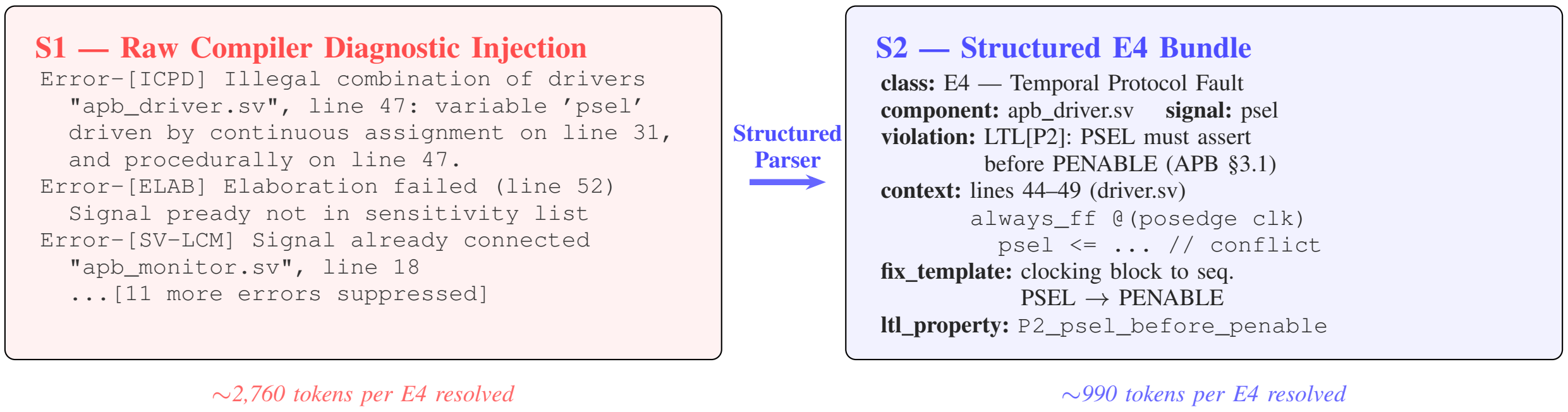

Analysis of generation attempts across the four protocol families
yielded six reproducible UVM error classes
(Table~\ref{tab:taxonomy}). These are the categories the
structured-feedback parser (S2) emits.
Fig.~\ref{fig:contrast} contrasts raw S1 injection with structured
S2 bundle injection for a representative E4 fault.

\begin{table}[t]
\caption{UVM Error Taxonomy}
\label{tab:taxonomy}
\centering
\small
\resizebox{\columnwidth}{!}{%
\begin{tabular}{@{}clcp{3.9cm}@{}}
\toprule
Class & Category & Freq. & Representative Symptom \\
\midrule
E1 & Factory registration   & 22\% & \texttt{create\_by\_name} returns null \\
E2 & TLM port/export mismatch & 19\% & connect type-check failure \\
E3 & Phase ordering violation & 17\% & null deref in \texttt{build\_phase} \\
E4 & Temporal protocol fault  & 24\% & SVA elab.\ / PREADY-PSEL violation \\
E5 & Scoreboard model error   & 11\% & burst beat-count mismatch \\
E6 & Parameterisation blindness & 7\% & hard-coded width elab.\ error \\
\bottomrule
\end{tabular}}
\end{table}

\subsection{Quality and Cost Metrics}
The \emph{Protocol Verification Completeness} (PVC) score
measures generated testbench quality as a weighted composite:
\begin{equation}
\text{PVC}(T) =
  \alpha\phi_{\text{struct}}
+ \beta\phi_{\text{proto}}
+ \gamma\phi_{\text{cov}}
+ \delta\phi_{\text{score}}
\label{eq:pvc}
\end{equation}
where $\phi_{\text{struct}}$, $\phi_{\text{proto}}$,
$\phi_{\text{cov}}$, and $\phi_{\text{score}}$ measure structural
completeness, protocol compliance, coverage intent alignment,
and scoreboard semantic correctness, each in $[0,1]$. Weights
$(\alpha,\beta,\gamma,\delta) = (0.20, 0.35, 0.25, 0.20)$ reflect
protocol-compliance primacy. Table~\ref{tab:sensitivity} reports
PVC under two alternative weightings; the
S0$<$S1$<$S2$<$S3 ordering is preserved in every case.

\begin{table}[t]
\caption{PVC Sensitivity to Weights (AHB2APB)}
\label{tab:sensitivity}
\centering
\small
\resizebox{\columnwidth}{!}{%
\begin{tabular}{@{}lccc@{}}
\toprule
Method & Baseline & Coverage-first & Structure-first \\
       & (0.20,0.35,0.25,0.20) & (0.15,0.25,0.40,0.20) & (0.35,0.30,0.20,0.15) \\
\midrule
S0 & 0.08 & 0.06 & 0.14 \\
S1 & 0.31 & 0.33 & 0.44 \\
S2 & 0.92 & 0.88 & 0.94 \\
S3 & 1.00 & 1.00 & 1.00 \\
\bottomrule
\end{tabular}}
\end{table}

The \emph{Protocol Complexity Index} (PCI) quantifies target
difficulty:
\begin{equation}
\text{PCI} = w_1|S| + w_2|\Sigma| + w_3 P_{\text{LTL}} + w_4 I_{\text{async}}
\label{eq:pci}
\end{equation}
where $|S|$ is FSM state count, $|\Sigma|$ the transaction
alphabet size, $P_{\text{LTL}}$ LTL property count,
$I_{\text{async}}$ asynchronous clock-domain crossings (0 or 1),
and $(w_1,w_2,w_3,w_4)=(1.0, 0.8, 1.2, 3.0)$.

Inference cost is reported as both total tokens per testbench
(a model-agnostic measure of inference work) and USD of API
spend at Claude Opus~4.8 published rates.

\subsection{Experimental Setup}
Four DUT targets: APB Subordinate (PCI\,=\,11.8), AHB-to-APB
bridge (PCI\,=\,14.2, primary benchmark), AHB DMA controller
(PCI\,=\,17.2), and AXI-Lite register slave (PCI\,=\,22.6).
Simulator: Mentor Questa~2021.2 with UVM-1.1d. Synthesis model:
Claude Opus~4.8. Per-cell results use $n=5$ independent seeds.
Every generated testbench is judged by the EDA toolchain
itself (\texttt{vlog} compile, \texttt{vsim} simulation, a
self-checking scoreboard, and array-based functional
coverage) rather than by model self-assessment.

\section{Results}
\label{sec:results}

\subsection{Cost-Quality Scaling Curve}
\label{sec:cost_quality}

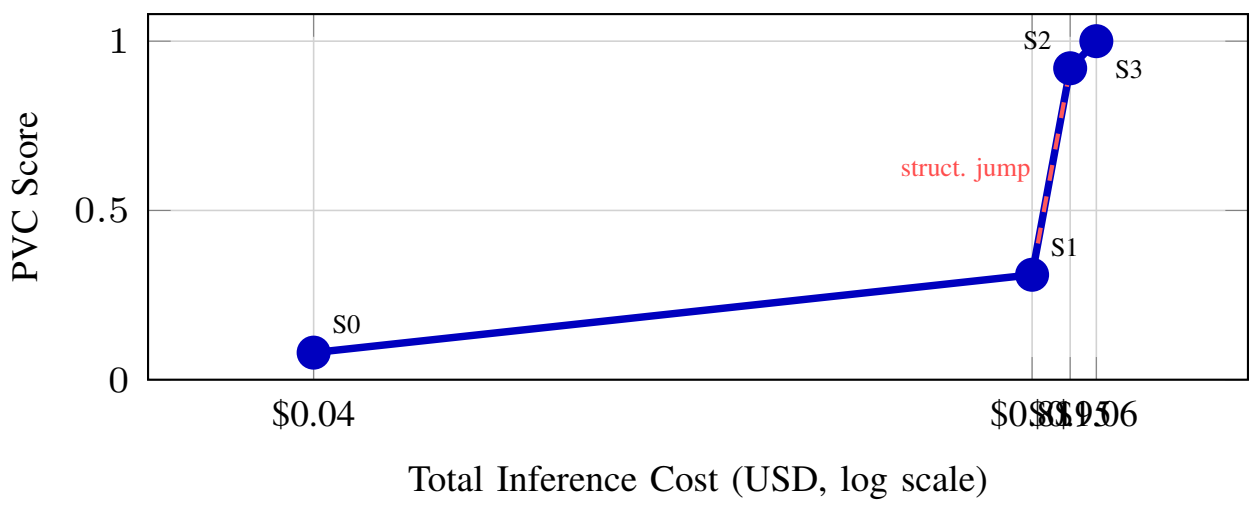
\begin{figure}[t]
\centering
\begin{tikzpicture}
\begin{axis}[
  width=\columnwidth,
  height=4.0cm,
  xlabel={Total Inference Cost (USD, log scale)},
  ylabel={PVC Score},
  xmode=log,
  xmin=0.02, xmax=2.0,
  ymin=0.0, ymax=1.08,
  xtick={0.04,0.81,0.95,1.06},
  xticklabels={\$0.04,\$0.81,\$0.95,\$1.06},
  ytick={0.0,0.5,1.0},
  grid=major,
  grid style={line width=0.3pt,draw=gray!35},
  tick label style={font=\scriptsize},
  label style={font=\scriptsize},
  every axis plot/.append style={line width=1.4pt},
]
\addplot[color=blue!75!black, mark=*, mark size=2.5pt]
  coordinates {(0.04,0.08)(0.81,0.31)(0.95,0.92)(1.06,1.00)};
\node[font=\tiny,above right] at (axis cs:0.04,0.08) {S0};
\node[font=\tiny,above right] at (axis cs:0.81,0.31) {S1};
\node[font=\tiny,above left]  at (axis cs:0.95,0.92) {S2};
\node[font=\tiny,below right] at (axis cs:1.06,1.00) {S3};
\draw[dashed,red!65,line width=0.8pt]
  (axis cs:0.81,0.31)--(axis cs:0.95,0.92);
\node[font=\tiny,red!70,left] at (axis cs:0.87,0.62)
  {struct.\ jump};
\end{axis}
\end{tikzpicture}
\caption{PVC vs.\ inference cost (log scale), AHB2APB. The
S1$\to$S2 jump (0.31$\to$0.92 for \$0.14) is the
structured-feedback compute multiplier.}
\label{fig:scaling}
\end{figure}

The return profile is asymmetric (Fig.~\ref{fig:scaling},
AHB2APB, $n=5$). S0$\to$S1
($\Delta\text{cost}\!\approx$\$0.77) yields
$\Delta\text{PVC}=+0.23$, a marginal efficiency of \textbf{0.30
PVC/\$}: raw feedback gets the testbench to compile, but with
the protocol fault still present the code compiles cleanly,
the compiler is silent, and quality stalls. S1$\to$S2
($\Delta\text{cost}\!\approx$\$0.14) yields
$\Delta\text{PVC}=+0.61$, efficiency \textbf{4.45 PVC/\$}; the
structured E1--E6 bundle identifies the protocol fault the
compiler cannot surface, and the fix lands. This is the
structured-feedback compute multiplier:
\textbf{$\sim$15$\times$} more PVC per dollar than raw feedback.
S2$\to$S3 ($\Delta\text{cost}\!\approx$\$0.11) yields
$\Delta\text{PVC}=+0.08$ (0.72 PVC/\$), driving functional
coverage from 69\% to 100\%.

Table~\ref{tab:main_results} reports the full grounded S0--S3
metrics. The per-stage marginal efficiencies (0.30, 4.45,
0.72~PVC/\$) make the multiplier transparent without relying
on a single aggregate ratio.

\begin{table}[t]
\caption{Multi-Strategy Results: AHB-to-APB (PCI=14.2), $n=5$}
\label{tab:main_results}
\centering
\small
\begin{tabular}{@{}lccccc@{}}
\toprule
Method & Compiles & PVC & Cov.\,(\%) & Tokens & Cost \\
\midrule
S0 Zero-shot     & No  & 0.08 & 0.0   & 2.7\,K  & \$0.04 \\
S1 Raw-feedback  & Yes & 0.31 & 43.8  & 120\,K  & \$0.81 \\
S2 Struct.-inner & Yes & 0.92 & 68.8  & 137\,K  & \$0.95 \\
S3 Full dual     & Yes & 1.00 & 100.0 & 153\,K  & \$1.06 \\
\bottomrule
\end{tabular}
\smallskip\\
\footnotesize PVC, compile, and coverage measured in Mentor
Questa; cost at Opus~4.8 rates. S0 does not compile;
S2/S3 pass the self-checking scoreboard with zero errors.
\end{table}

\subsection{Mechanism: Fault Targeting vs.\ Token Reduction}
\label{sec:tokens}
The structured inner loop is disproportionately cost-efficient
not because it reduces raw token throughput (S2 and S1 consume
comparable tokens overall, Table~\ref{tab:main_results}), but
because it resolves a class of faults that raw feedback cannot.
The structured bundle (error class, component, code context,
violated LTL property, fix template) classifies and localises
the functional faults (E4, E5) that leave the code syntactically
valid, directing inference compute at the failing constraint.
Fig.~\ref{fig:contrast} illustrates this contrast for a
representative E4 fault.

\subsection{Verification Grounding and Bug Discovery}
\label{sec:bugfind}
The dual-loop (S3) testbenches on APB-SLV and AXI-Lite produced
repeatable scoreboard failures (42 and 45 \texttt{UVM\_ERROR}s
respectively), all of the same form: an out-of-range protocol
error response on accesses that should have succeeded. Tracing
to the RTL revealed a genuine design bug in both
subordinates: an address-decode comparison
\texttt{addr[\,$\cdot$\,]\,<\,NUM\_REGS[\,$\cdot$\,]} in which
the parameter was bit-truncated, so the in-range predicate was
always false. The generated testbenches thus functioned as
working verification IP and surfaced a latent RTL bug, an outcome
outside the reach of LLM-judged metrics that score code which
never executed against a DUT. Against
corrected RTL, all four S3 testbenches pass with zero scoreboard
errors (Table~\ref{tab:multi_protocol}).

\subsection{Multi-Protocol Results}
\label{sec:pci}
Table~\ref{tab:multi_protocol} reports grounded S3 metrics on
all four targets. The dual-loop strategy reaches
PVC\,$\geq$\,0.95 across PCI\,11.8--22.6, with every target
passing its self-checking scoreboard at zero errors. Coverage
closure differs by bin cross-product size: 100\% on APB-SLV and
AHB-DMA; 81\% on AHB2APB and 91\% on AXI-Lite under the same
iteration budget.

\begin{table}[t]
\caption{Multi-Protocol Results: Strategy S3 (grounded, $n=5$)}
\label{tab:multi_protocol}
\centering
\small
\begin{tabular}{@{}lcccc@{}}
\toprule
Target & PCI & PVC & Cov.\,(\%) & Scoreboard \\
\midrule
APB-SLV  & 11.8 & 1.00 & 100.0 & pass (5/5) \\
AHB2APB  & 14.2 & 0.95 & 81.3  & pass (5/5) \\
AHB-DMA  & 17.2 & 1.00 & 100.0 & pass (5/5) \\
AXI-Lite & 22.6 & 0.98 & 91.1  & pass (5/5) \\
\bottomrule
\end{tabular}
\end{table}

\section{Discussion}
\label{sec:discussion}
\textbf{Compute multiplier mechanism.} The $\sim$15$\times$
per-dollar advantage reflects a specific mechanism: raw feedback
addresses what the compiler reports, but the dominant faults
(E4 temporal, E5 scoreboard) leave the code compilable and
produce no compiler diagnostic; the structured bundle localises
those functional faults directly.

\textbf{Industrial implications.} At \$1.06 per S3 testbench,
agentic synthesis costs a small fraction of a DV engineer-hour;
toolchain grounding, not model self-assessment, is what lets the
generated environments function as working verification IP, as
the latent RTL bug they surfaced shows.

\textbf{Limitations.} PCI and PVC weights are expert-assigned;
the S0$<$S1$<$S2$<$S3 ordering is invariant under three
weightings (Table~\ref{tab:sensitivity}). The evaluation uses one
model (Claude Opus~4.8) and four AMBA targets; open-weight or
SystemVerilog-tuned models, multi-CDC designs, and non-protocol
verification remain open. Coverage is measured against
manifest-declared bins; structural code coverage would require
a coverage-licensed simulator.

\section{Conclusion}
\label{sec:conclusion}
We have shown that structured diagnostic feedback acts as an
inference-time compute multiplier for agentic UVM synthesis:
the structured inner loop yields $\sim$15$\times$ more PVC per
dollar than raw feedback by targeting functional faults that
leave the code compilable. Across four AMBA targets
(PCI\,11.8--22.6), the dual-loop strategy reaches
PVC\,$\geq$\,0.95 with zero scoreboard errors at $\sim$\$1 per
testbench while surfacing a latent RTL bug in two of the four
reference designs.


\end{document}